\documentclass[3p,a4paper,twocolumn]{elsarticle}

\usepackage{amsfonts}
\usepackage{amsmath}
\usepackage{amssymb}
\usepackage{graphicx}
\begin{document}

\begin{frontmatter}

\title{Symmetry aspects of quantum discord}

\author[jm]{J. Maziero\corref{cor1}\fnref{fn1}}
\ead{jonasmaziero@gmail.com}

\author[jm]{L. C. C\'{e}leri}
\ead{lucas.celeri@ufabc.edu.br}

\author[jm]{R. M. Serra}
\ead{serra@ufabc.edu.br}

\address[jm]{Centro de Ci\^{e}ncias Naturais e Humanas, Universidade Federal do ABC, R.
Santa Ad\'{e}lia 166, 09210-170, Santo Andr\'{e}, S\~{a}o Paulo, Brazil.}



\begin{abstract}
Quantum discord (QD) reveals the nonclassical nature of correlations in bipartite quantum states, going beyond the entanglement-separability paradigm. In this article we discuss the suitability of QD in what concern its possible asymmetry with relation to the bipartition we choose to compute it. We obtain an analytical formula for a symmetrized version of QD (SQD) in Bell-diagonal states. We observe that if correlation is regarded as a shared property, then the SQD could be a convenient quantifier for asymmetric states.
\end{abstract}


\end{frontmatter}


\section{Introduction}

The quantumness of correlations in composed quantum systems is believed to be the resource behind the advantage of protocols developed (to process and transmit information) in quantum information theory (QIT) over its classical counterparts. Therefore one of the main problems in QIT is the characterization and quantification of such a resource. The total correlation in a bipartite quantum state $\rho_{AB}$ is measured by the quantum mutual information (QMI) \cite{Schumacher} $\mathcal{I}\left(  \rho_{AB}\right)  =S\left(  \rho_{A}\right)  +S\left(  \rho_{B}\right)  -S\left(\rho_{AB}\right)  $, where $S\left(  \rho\right)  =-\operatorname*{Tr}\left(\rho\log_{2}\rho\right)  $ is the von Neumann entropy. Thus, an important related problem is how to separate the total correlation in its classical and quantum contributions \cite{Henderson}.

Until some time ago, entanglement (a correlation associated with the separability problem \cite{Horodecki1}) was usually regarded as the only kind of nonclassical correlation in a composed state. However, while this idea is valid for bipartite pure states it is not necessarily true for mixed ones \cite{Ollivier,Oppenheim,Vedral}. This last class of states can support quantum correlation other than entanglement, e.g., the correlation revealed by the so-called quantum discord (QD) \cite{Ollivier}. Moreover, this general nonclassical correlation can be present even in separable states \cite{Ollivier} and may, in principle, be useful for applications in quantum computing \cite{Datta1}.

Recently, the dynamic behavior of classical and nonclassical correlations under decoherence have shown to be very peculiar \cite{Maziero1, Maziero2}. The decay rates of such correlations may exhibit a sudden change in behavior \cite{Maziero1}, which was experimentally observed in an optical setup in Ref. \cite{Xu}. In some situations, in which a system immersed in an environment that does not exchange energy with the former, the decoherence phenomenon may occur without entanglement between the system and reservoir \cite{Maziero2}. This result contradicts our intuition about the cause of the relaxation process. In such a case the mechanism for the unavoidable information transfer from the system to the environment, and therefore for the coherence loss, is provided by separable-state quantum correlation and its classical counterpart \cite{Maziero2}. It suggests that besides technological applications, quantifying the quantumness of correlations in separable states is also important from a fundamental point of view.

Our intention in this article is to study the symmetry aspects of some quantum correlation quantifiers. It is well known that the value of QD can be different depending on the bipartition we choose to compute it. We should mention that in recent years the possible asymmetric quantifier QD have been employed in several scenarios, e.g., quantum phase transitions \cite{Dillenschneider1}, open quantum systems \cite{Maziero1,Maziero2,Rosario}, biological systems \cite{Bradler}, thermodynamics \cite{Oppenheim,Horodecki2}, relativistic effects in QIT \cite{Datta2}, and so on. Thus, it is important to determine in what situations the QD can be used as an adequate quantifier, and when more suitable measures of correlations should be adopted. As already expected, when symmetric definitions for correlations are used, we have a price to pay regarding its computability through an optimization process which involves more degrees of freedom. In this connection we obtain, in Section 3, an analytical expression for a symmetric definition of classical correlation and its quantum counterpart [a symmetric quantum discord (SQD)] for the so-called two-qubit Bell-diagonal states. We also present, in Section 4, explicit examples of classes of states where the asymmetrical quantifiers may be not convenient since different values for the correlations are obtained depending on the measurement process adopted.


\section{Correlation measures}

In Ref. \cite{Henderson} the authors proposed a quantifier for classical correlation based on the reduction of uncertainty in the state of one partition of a bipartite system, due to measurements performed on the other partition, possibly correlated with the former. This quantifier can be written as
\begin{equation}
\mathcal{C}_{Y}\left(  \rho_{AB}\right)=S\left(  \rho_{X}\right)-\min_{\left\{\Pi_{j}^{Y}\right\}}S_{\Pi}\left(  \rho_{\left. X\right\vert Y}\right)  \text{,} 
\label{CC}
\end{equation}
where the quantum conditional entropy of $X$ when the partition $Y$ is measured is given by $S_{\Pi}\left(  \rho_{\left.  X\right\vert Y}\right)={\textstyle\sum_{j}}p_{j}^{X}S\left(  \rho_{j}^{X}\right)  $. The state $\rho_{j}^{X}=\left.\operatorname*{Tr}\nolimits_{Y}\left(  \Pi_{j}^{Y}\rho_{AB}\Pi_{j}^{Y}\right)\right/  p_{j}^{Y}$ is the post measurement reduced density operator of partition $X$ which is obtained with probability $p_{j}^{X}=\operatorname*{Tr}\left(\Pi_{j}^{Y}\rho_{AB}\right)  $. The measurement on partition $Y$ that reaches the minimum in Eq. (\ref{CC}) is that yielding the maximum information about the state of the partition $X$. The pair $X,Y$ can be chosen to be $A,B$ or $B,A$. Here, the minimum is taken over a complete set of orthogonal projective measurements $\left\{\Pi_{j}^{Y}\right\}  $ on the partition $Y$ instead of the more general positive operator-valued measurements (POVM) used in the original definition \cite{Henderson}. It was shown, in Ref. \cite{Hamieh}, that for two-qubit states Eq. (\ref{CC}) is optimized through rank-1 POVMs. In several practical applications, the quantity in Eq. (\ref{CC}) is often computed (in a simplified way) considering orthogonal projective measurements. Such a procedure is largely adopted since it gives, at least, a lower bound for the classical content of correlation. So, we can think of (\ref{CC}) as quantity related to the classical nature of correlations (depending on the state considered it could be the classical correlation itself or just a bound). Here we will adopt quantities optimized using orthogonal projective measurements for the sake of addressing the issue of symmetry and asymmetry of correlations. Nevertheless, the analysis of such issues using general POVMs is an interesting problem for future investigations.

The quantum correlation may be defined as
\begin{equation}
\mathcal{Q}_{Y}\left(  \rho_{AB}\right)  =\mathcal{I}\left(  \rho_{AB}\right) -\mathcal{C}_{Y}\left(\rho_{AB}\right)  \text{,}
\label{QC}
\end{equation}
which, under the conditions regarded here, is identical to the definition of quantum discord introduced in Ref. \cite{Ollivier}. So, Eq. (\ref{CC}) is the classical counterpart of the quantum discord and \textit{vice versa} \cite{Maziero1,Hamieh,Luo1}. In the particular case of pure states, the QD is equal to the entropy of entanglement and also equal to the classical correlation in Eq. (\ref{CC}), i.e., $\mathcal{Q}_{Y}\left(\rho_{AB}\right) =\mathcal{C}_{Y}\left(  \rho_{AB}\right)  =\left.  \mathcal{I}(\rho_{AB})\right/  2$ \cite{Henderson}, independent of the choice of the partition $Y$. On the other hand, for mixed states the entanglement is only a part of this general nonclassical correlation \cite{Ollivier}. Indeed, the QD may be distributed into entanglement and an additional quantum
correlation as proposed in Ref. \cite{Modi}.

Recently, some symmetric measures of correlations were proposed \cite{Oppenheim,Modi,Terhal,Piani}. Alternatively, the classical correlation in a composed bipartite system may be defined as the maximum mutual information obtained via a complete set of local von Neumann measurements over both partitions of a composite state $\rho_{AB}$:
\begin{equation}
\mathcal{C}_{S}\left(\rho_{AB}\right)=\max_{\Pi^{AB}}I(\Pi^{AB}[\rho_{AB}]) \text{,}\label{sCC}
\end{equation}
where the local projective-measurement map, $\Pi^{AB}[\rho_{AB}]$, (LPMM) is explicitly given by 
\begin{equation*}
\Pi^{AB}[\rho_{AB}]=\sum_{i,j}\Pi_{i}^{A}\otimes\Pi_{j}^{B}(\rho_{AB})\Pi_{i}^{A}\otimes\Pi_{j}^{B}\text{,} 
\end{equation*}
with $\sum_{i}\Pi_{i}^{A(B)}=\mathbf{1}^{A(B)}$ and $\Pi_{i}^{A(B)}\Pi_{j}^{A(B)}=\delta_{ij}\Pi_{i}^{A(B)}$. We define the quantum counterpart of Eq. (\ref{sCC}) as
\begin{equation}
\mathcal{Q}_{S}\left(\rho_{AB}\right)  =\mathcal{I}(\rho_{AB})-\mathcal{C}
_{S}(\rho_{AB})\text{.} \label{sQC}
\end{equation}
The quantum correlation defined in Eq. (\ref{sQC}) can be seen as a symmetric version of the quantum discord (SQD), since it also reveals a departure between the quantum and classical versions of information theory, i.e., the different nature of the measurement process. Such a symmetrized version of the quantum discord was recently employed in an experiment to investigate the quantumness of correlations in a  spin system, at room temperature, in a nuclear magnetic resonance setup \cite{NMR}. 

It should be noted that both QD and SQD, as defined here, are not monotonic under local quantum operations (LO). In fact, such quantities should be understood as upper bounds for the quantum nature of correlations.  We observe also that, as mutual information is monotonic under LO \cite{Nielsen}, it follows $\mathcal{Q}_{S}\left(\rho_{AB}\right)\ge 0$. Moreover,  we prove bellow (following \cite{Piani}) that $\mathcal{Q}_{S}\left(\rho_{AB}\right)=0$ if and only if the state $\rho_{AB}$ is the incorporation of a classical probability distribution in the quantum formalism, i.e., the so called classical-classical state \cite{Piani}.

\textit{Theorem.} $\mathcal{Q}_{S}\left(\rho_{AB}\right)=0$ if and only if the state $\rho_{AB}$ can be cast in the form $\sum_{i,j}p_{i,j}\Pi_{i}^{A}\otimes\Pi_{j}^{B}$. (Here $\{\Pi_{i}^{A(B)}\}$ is a complete set of projectors for the subsystem $A(B)$ and $p_{i,j}$ is a probability distribution.)

\textit{Proof.} First we assume $\rho_{AB}=\sum_{i,j}p_{i,j}\Pi_{i}^{A}\otimes\Pi_{j}^{B}$.
If we apply the LPMM in the eigenbasis of $\rho_{AB}$ it remains undisturbed and, as a consequence, $\max_{\Pi^{AB}}I(\Pi^{AB}[\rho_{AB}])=I(\{p_{i,j}\})=I(\rho_{AB})$. Thus  $\mathcal{Q}_{S}\left(\rho_{AB}\right)=0$. Now we assume $\mathcal{Q}_{S}\left(\rho_{AB}\right)=0$, which implies $I(\rho_{AB})=\max_{\Pi^{AB}}I(\Pi^{AB}[\rho_{AB}])$. If $\tilde{\Pi}^{AB}$ is taken as the LPMM attaining the maximum in the right hand side of the last equation, then $I(\rho_{AB}) = I(\tilde{\Pi}^{AB}[\rho_{AB}])$. Using the Petz's theorem (see \cite{Petz} and \textit{Lemma 2} in Ref. \cite{Piani}), we see that the last equation implies the existence of LO $\Lambda^{A}$ and $\Lambda^{B}$ such that $\rho_{AB}=\Lambda^{A}\otimes\Lambda^{B}[\tilde{\Pi}^{AB}[\rho_{AB}]]=\sum_{i,j}\tilde{p}_{ij}\Lambda^{A}[\tilde{\Pi}_{i}^{A}]\otimes\Lambda^{B}[\tilde{\Pi}_{j}^{B}]$. Let us now consider the following state: $\rho^{cq}_{AB}=\mathcal{\mathcal{I}}^{A}\otimes\Lambda^{B}[\tilde{\Pi}^{AB}[\rho_{AB}]]=\sum_{i}\tilde{p}_{i}\tilde{\Pi}_{i}^{A}\otimes\sum_{j}\frac{\tilde{p}_{ij}}{\tilde{p}_{i}}\Lambda^{B}[\tilde{\Pi}_{j}^{B}]=\sum_{i}\tilde{p}_{i}\tilde{\Pi}_{i}^{A}\otimes\tilde{\rho}_{i}^{B}$. 
As mutual information does not increases under LO, we have $I(\tilde{\Pi}^{AB}[\rho_{AB}])\ge I(\rho^{cq}_{AB})$. But $\tilde{\Pi}^{AB}[\rho_{AB}]=\sum_{i,j}\tilde{p}_{ij}\tilde{\Pi}_{i}^{A}\otimes\tilde{\Pi}_{j}^{B}$ can be obtained from $\rho^{cq}_{AB}$ through a LO applied to the subsystem $B$, and therefore $I(\tilde{\Pi}^{AB}[\rho_{AB}])\le I(\rho^{cq}_{AB})$, which leads to $I(\tilde{\Pi}^{AB}[\rho_{AB}])=I(\rho^{cq}_{AB})$. However, it is known from the Holevo's bound \cite{Nielsen}, that $I(\rho^{cq}_{AB})=\chi(\{\tilde{p}_{i},\tilde{\rho}_{i}^{B}\})=I(\tilde{\Pi}^{AB}[\rho_{AB}])$
if and only if all $\tilde{\rho}_{i}^{B}$ commute, where $\chi(\{\tilde{p}_{i},\tilde{\rho}_{i}^{B}\})=S(\sum_{i}\tilde{p}_{i}\tilde{\rho}_{i}^{B})-\sum_{i}\tilde{p}_{i}S(\tilde{\rho}_{i}^{B})$ is the Holevo's quantity. Therefore $\rho^{cq}_{AB}$ is of the form $\sum_{i,j}p_{i,j}\Pi_{i}^{A}\otimes\Pi_{j}^{B}$. Finally, by defining $\rho_{AB}=\Lambda^{A}\otimes\mathcal{\mathcal{I}}^{B}[\rho^{cq}_{AB}]$
and using the same reasoning as above we see that $\rho_{AB}$ also must take the form $\sum_{i,j}p_{i,j}\Pi_{i}^{A}\otimes\Pi_{j}^{B}$, completing the proof.


\section{Symmetric quantum discord for Bell-diagonal states}

Any two-qubit state can be brought through local unitaries to the following form:
\begin{align}
\rho_{AB}  &  =\frac{1}{4}\{\mathbf{1}^{AB}+{\sum_{k=1}^{3}}
c_{k}\sigma_{k}^{A}\otimes\sigma_{k}^{B}+\mathbf{a}\cdot\mathbf{\sigma}
^{A}\otimes\mathbf{1}^{B}\nonumber\\
&  \text{ \ \ \ \ \ \ }+\mathbf{1}^{A}\otimes\mathbf{b}\cdot\mathbf{\sigma
}^{B}\}\text{,} \label{2qubits}
\end{align}
where $c_{k}\in
\mathbb{R}
$, $\mathbf{a,b\in
\mathbb{R}
}^{3}$, and $\mathbf{\sigma}^{X}=\left(  \sigma_{1}^{X},\sigma_{2}^{X},\sigma_{3}^{X}\right)  $ with $\sigma_{k}^{X}$ being the $k$th component of the Pauli operator acting on the state space of partition $X=A,B$. The
identity operator acting on the state space $\mathcal{E}_{AB}$, $\mathcal{E}_{A}$ and $\mathcal{E}_{B}$ was defined as\textbf{ }$\mathbf{1}^{AB}$, $\mathbf{1}^{A}$ and $\mathbf{1}^{B}$, respectively.

Let us first compute the symmetric classical (\ref{sCC}) and quantum (\ref{sQC}) correlations regarding the class of two-qubit states with maximal mixed marginals (Bell-diagonal states) $\rho_{AB}^{m}$. These states are obtained by choosing $\mathbf{a}=\mathbf{b}=\mathbf{0}$ in Eq. (\ref{2qubits}). Now the coefficients $c_{k}$ are constrained such that the eigenvalues of $\rho_{AB}^{m}$, $\lambda_{i,j}=\left[  1+(-1)^{i}c_{1}-(-1)^{i+j}
c_{2}+(-1)^{j}c_{3}\right] /4$ ($i,j=0,1$), are not negative. Since $S\left(\rho_{A}^{m}\right)  =S\left(  \rho_{B}^{m}\right) =1$ the total correlation in this class of states is given by $\mathcal{I}\left(\rho_{AB}^{m}\right)=2+{\textstyle\sum\nolimits_{i,j=0}^{1}}\lambda_{i,j}\log_{2}\lambda_{i,j}$.

If the system is prepared in the state $\rho_{AB}^{m}$ the two-side measured density operator reads
\begin{equation}
\label{Mmmm}
\eta_{AB}^{m}=\frac{\mathbf{1}^{AB}}{4}+{\sum_{i,j=0}^{1}}{\sum_{k=1}^{3}}\frac{c_{k}}{4}\Pi_{i}^{A}\sigma_{k}^{A}\Pi_{i}^{A}\otimes\Pi_{j}^{B}\sigma_{k}^{B}\Pi_{j}^{B}\text{,}
\end{equation}
and the reduced measured density operator is $\eta_{A(B)}^{m}=\left. \mathbf{1}^{A(B)}\right/  2$. Thus $S\left(\eta_{A}^{m}\right) =S\left(\eta_{B}^{m}\right) =1$ and the symmetric classical correlation (\ref{sCC})
can be written as
\begin{equation*}
 C_{S}\left( \rho_{AB}^{m}\right)  =2-\min\limits_{\left\{\mathbf{M}_{i,j}\right\}  }S\left(  \eta_{AB}^{m}\right).
\end{equation*}
Following \cite{Luo1}, any projective measurement can be obtained through a $U(2)$ rotation $R_{Y}$ on a projector $\left\vert j_{Y}\right.  \rangle\langle\left.  j_{Y}\right\vert $ as
\begin{equation}
\Pi_{j}^{Y}=R_{Y}\left\vert j_{Y}\right.  \rangle\langle\left.  j_{Y}
\right\vert R_{Y}^{\dagger}\text{,} \label{MO}
\end{equation}
with $j_{Y}=0_{Y},1_{Y}$. Here $\left\{  \left\vert 0_{Y}\right.\rangle,\left\vert 1_{Y}\right.  \rangle\right\}  $ is the standard computational basis for the partition $Y$\textbf{, }with\textbf{ }$Y=A,B$. By using the fact that $\left\{  \mathbf{1}^{Y},\sigma_{1}^{Y},\sigma_{2}^{Y},\sigma_{3}^{Y}\right\}  $ are the generators for the $SU(2)$ group we can write $R_{Y}=u_{0}^{Y}\mathbf{1}^{Y}+i\mathbf{u}^{Y}\cdot\mathbf{\sigma}^{Y}$, where $\mathbf{u}^{Y}=\left(  u_{1}^{Y},u_{2}^{Y},u_{3}^{Y}\right)  $ with $u_{l}^{Y}$ $\mathbf{\in\mathbb{R}}$, $l=0,1,2,3$. The unitarity of $R_{Y}$ implies that $\left(  u_{0}^{Y}\right)  ^{2}+\left(  u_{1}^{Y}\right)  ^{2}+\left(  u_{2}^{Y}\right)^{2}+\left(  u_{3}^{Y}\right)  ^{2}=1$. Using this definition for $R_{Y}$, it is straightforward to show that
\begin{equation}
{\sum_{j=0}^{1}}\Pi_{j}^{Y}\sigma_{k}^{Y}\Pi_{j}^{Y}=v_{k}^{Y}R_{Y}\sigma_{3}^{Y}
R_{Y}^{\dagger}\text{,} \label{sum}
\end{equation}
where we set $v_{1}^{Y}\equiv2\left(  u_{1}^{Y}u_{3}^{Y}-u_{0}^{Y}u_{2}^{Y}\right)  $, $v_{2}^{Y}\equiv2\left(u_{2}^{Y}u_{3}^{Y}+u_{0}^{Y}u_{1}^{Y}\right)  $, and $v_{3}^{Y}\equiv(u_{0}^{Y})^{2}-(u_{1}^{Y})^{2}-(u_{2}^{Y})^{2}+(u_{3}^{Y})^{2}$. It can be verified that $\left(  v_{1}^{Y}\right)^{2}+\left(  v_{2}^{Y}\right)  ^{2}+\left(  v_{3}^{Y}\right)  ^{2}=1$. If we define $w\equiv{\textstyle\sum\nolimits_{i=1}^{3}} c_{i}v_{i}^{A}v_{i}^{B}$ and replace Eq. (\ref{sum}) in Eq. (\ref{Mmmm}) the measured density operator reads $\eta_{AB}^{m}=\left(  \mathbf{1}^{AB}+wR_{A}\sigma_{3}^{A}R_{A}^{\dagger}\otimes R_{B}\sigma_{3}^{B}R_{B}^{\dagger}\right)/4$ and has the following eigenvalues spectrum $\lambda_{i}^{m}=\left.  \left[  1+(-1)^{i}w\right]  \right/  4$, with $i=1,2,3,4$. The minimization of the measured density operator entropy, $S\left(  \eta_{AB}^{m}\right)=-\sum_{i=1}^{4}\lambda_{i}^{m}\log_{2}\lambda_{i}^{m}$, is obtained by maximizing the absolute value of $w$. For this purpose it is helpful to note that
\begin{align*}
\left\vert w\right\vert  &  \leq\left\vert c_{1}\right\vert \left\vert v_{1}^{A}v_{1}^{B}\right\vert +\left\vert c_{2}\right\vert \left\vert v_{2}^{A}v_{2}^{B}\right\vert +\left\vert c_{3}\right\vert \left\vert v_{3}^{A}v_{3}^{B}\right\vert \\
&  \leq\kappa\left(  \left\vert v_{1}^{A}v_{1}^{B}\right\vert +\left\vert
v_{2}^{A}v_{2}^{B}\right\vert +\left\vert v_{3}^{A}v_{3}^{B}\right\vert
\right)  \leq\kappa,
\end{align*}
where we defined $\kappa\equiv\max\left(  \left\vert c_{1}\right\vert,\left\vert c_{2}\right\vert ,\left\vert c_{3}\right\vert \right)  $. The last inequality is saturated to an equality if the measurement directions are the
same in both partitions $A$ and $B$, i.e. $v_{1}^{A}=v_{1}^{B}$, $v_{2}^{A}=v_{2}^{B}$ and $v_{3}^{A}=v_{3}^{B}$. Thus the maximum of $\left\vert w\right\vert $ is obtained choosing $u_{l}^{Y}$ appropriately. For example, if $\kappa=\left\vert c_{1}\right\vert $ we have $v_{1}^{Y}=1$ and $v_{2}
^{Y}=v_{3}^{Y}=0$. This condition is fulfilled with $u_{1}^{Y}=u_{3}^{Y}=1$ and $u_{0}^{Y}=u_{2}^{Y}=0$ or with $u_{1}^{Y}=u_{3}^{Y}=0$ and $u_{0}^{Y}=u_{2}^{Y}=1$. The other alternatives, $\kappa=\left\vert c_{2}\right\vert$ and $\kappa=\left\vert c_{3}\right\vert $, are obtained using the same reasoning. Thus the symmetric classical correlation (\ref{sCC}) in the state $\rho_{AB}^{m}$ can be analytically obtained and it is given by
\begin{equation}
\mathcal{C}_{S}\left(  \rho_{AB}^{m}\right)={\sum_{i=0}^{1}}
\frac{1+(-1)^{i}\kappa}{2}\log_{2}[1+(-1)^{i}\kappa]\text{.} \label{Acc}
\end{equation}
The expression (\ref{Acc}) for the symmetric classical correlation (\ref{sCC}) of two-qubit states with maximal mixed marginals, $\rho_{AB}^{m}$, coincide with the expression obtained by Luo \cite{Luo1} for the asymmetric classical correlation (\ref{CC}) in this case. Therefore, for this class of states, the SQD $\mathcal{Q}_{S}(\rho_{AB}^{m})=\mathcal{I}(\rho_{AB}^{m})-\mathcal{C}_{S}\left(  \rho_{AB}^{m}\right)  $ reduces to the asymmetric QD, in other words $\mathcal{Q}_{S}(\rho_{AB}^{m})=\mathcal{Q}_{Y}(\rho_{AB}^{m})$ for $Y=A$ or $B$.


\section{Asymmetry of quantum discord exemplified}

For the sake of introducing explicit examples of classes of states for which the quantum discord may not be a convenient quantifier of quantum correlation, we use the X-real class of states $\rho_{AB}^{x}$. These states are obtained from (\ref{2qubits}) by choosing $a_{1}=$ $a_{2}=b_{1}=b_{2}=0$ and have the following spectrum of eigenvalues: $\lambda_{i,j}^{x}=[1+(-1)^{j}c_{3}+(-1)^{i}\sqrt{(c_{1}-(-1)^{j}c_{2})^{2}+(a_{3}+(-1)^{j}b_{3})^{2}}]/4$, with $i,j=0,1$. The reduced state entropy is given by
\[
S(\rho_{X}^{x})=-\sum_{i=0}^{1}\{[1+(-1)^{i}\varsigma_{3}]/2\}\log
_{2}\{[1+(-1)^{i}\varsigma_{3}]/2\},
\]
where $\varsigma_{3}=a_{3}(b_{3})$ for $X=A(B)$. Thus the total correlation in $\rho_{AB}^{x}$ is $\mathcal{I}(\rho_{AB}^{x})=S(\rho_{A}^{x})+S(\rho_{B}^{x})+{\textstyle\sum\nolimits_{i,j=0}^{1}}\lambda_{i,j}^{x}\log_{2}\lambda_{i,j}^{x}$. In Ref. \cite{Ali} the authors obtained an analytical expression for asymmetric classical correlation [Eq. (\ref{CC})], and thus for the QD, in X-real states. This expression is given by $\mathcal{C}_{Y}(\rho_{AB}^{x})=S(\rho_{X}^{x})-\min[S_{\Pi}^{(1)}(\rho_{X\left\vert Y\right.  }^{x})$,$S_{\Pi}^{(2)}(\rho_{X\left\vert Y\right.  }^{x})]$, where the conditional entropies are given by $S_{\Pi}^{(1)}(\rho_{X\left\vert Y\right.}^{x})=S[\rho_{AB}^{x}(c_{1}=c_{2}=0)]-S(\rho_{Y}^{x})$ and $S_{\Pi}^{(2)}(\rho_{X\left\vert Y\right.}^{x})=-\sum_{i=0}^{1}\{[1+(-1)^{i}\sqrt{\varsigma_{3}^{2}+c^{2}}]/2\}\log_{2}\{[1+(-1)^{i}\sqrt{\varsigma_{3}^{2}+c^{2}}]/2\}$, with $X=A$, $Y=B$ or $X=B$, $Y=A$ and $c\equiv\max(|c_{1}|,|c_{2}|)$.
\begin{figure}[!]
\begin{center}
\includegraphics[height=2.3in,width=3.0in]{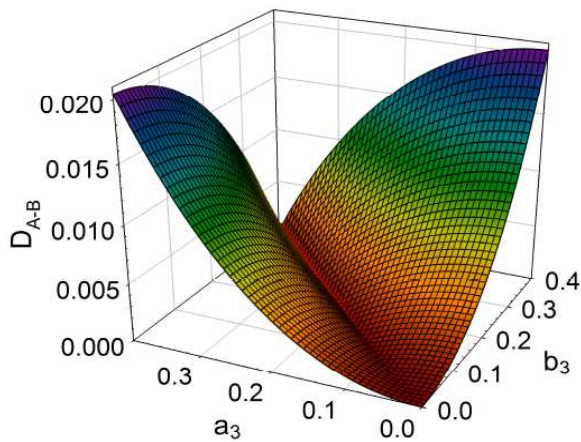}
\caption{(Color online) Absolute value of the difference between the quantum
discord in $\rho_{AB}^{x}$ obtained by measuring $A$ and $B$, $D_{A-B}$,\ as a
function of $a_{3}$ and $b_{3}$ for $c_{1}=0.6$, $c_{2}=0$ (or $c_{1}=0$,
$c_{2}=0.6$), and $c_{3}=0$.}
\label{fig1}
\end{center}
\end{figure}
We note then that if $a_{3}=b_{3}$ we have $S(\rho_{A}^{x})=S(\rho_{B}^{x})$ and thus $\mathcal{Q}_{B}(\rho_{AB}^{x})=\mathcal{Q}_{A}(\rho_{AB}^{x}$). However for $a_{3}\neq b_{3}$ we can obtain $\mathcal{C}_{B}(\rho_{AB}^{x})\neq\mathcal{C}_{A}(\rho_{AB}^{x}$) depending on the state regarded. For
instance, if $c_{1}=0.6$, $c_{2}=0$ (or equivalently for $c_{1}=0$, $c_{2}=0.6$), and $c_{3}=0$ the QD obtained by measuring $A$ ($\mathcal{Q}_{A}$) or $B$ ($\mathcal{Q}_{B}$) always yields different values. The absolute
value of the difference between the two quantifiers, $D_{A-B}\equiv\left\vert \mathcal{Q}_{A}(\rho_{AB}^{x})-\mathcal{Q}_{B}(\rho_{AB}^{x})\right\vert $, is plotted in Fig. 1. The extreme situation is obtained for $a_{3}=0$ and $b_{3}\neq0$ (or $b_{3}=0$ and $a_{3}\neq0$). In this case the measurement in one partition results in a zero value for the QD while a nonzero value is obtained by measuring the other partition. Thus, if we regard the correlation between the two qubits as a shared property, QD may be not a convenient quantifier for quantum correlations in this case, although it still can be used as an indicator in a proper way  (we have to compute the two asymmetric versions of QD). 


\section{Conclusions}

In summary, we briefly analyzed some symmetric and asymmetric measures of correlations. Considering a two-qubit composed system, we obtained an analytical formula for a symmetric version of quantum discord for Bell-diagonal states. We show that in this case the SQD and QD are equivalent. We also pointed out the fact that, if correlation is regarded as a shared property, quantum discord may be not convenient to quantify quantum correlation in a general scenario due to its asymmetry with respect to the chosen partition to be measured. As illustrated in Sec. 4, such asymmetry can leads us to find different values for the amount of correlations shared between the two qubits, depending on the choice of the measured partition. In this case, the asymmetric quantum discord may be regarded as an indicator of nonclassical correlations, but an appropriate strategy to compute it must be adopted. We then observe that a symmetrized version of quantum discord could be more convenient to quantify quantum correlations in general scenarios. An important open problem, which must be addressed in the future, is related with the investigation of the optimization process in SQD through a generic POVM.  


\section*{Acknowledgments}

We are grateful for the funding from UFABC, CAPES, and FAPESP. This work was performed as part of the Brazilian National Institute of Science and Technology for Quantum Information (INCT-IQ).

\bibliographystyle{elsarticle-num}

\begin{thebibliography}{00}                                                                                        

\bibitem{Schumacher} B. Schumacher and M. D. Westmoreland, Phys. Rev. A 74 (2006) 042305; B. Groisman, S. Popescu, and A. Winter, Phys. Rev. A 72 (2005) 032317.

\bibitem {Henderson}L. Henderson and V. Vedral, J. Phys. A: Math. Gen. 34 (2001) 6899.

\bibitem {Horodecki1}R. Horodecki, P. Horodecki, M. Horodecki, and K. Horodecki, Rev. Mod. Phys. 81 (2009) 865.

\bibitem {Ollivier}H. Ollivier and W. H. Zurek, Phys. Rev. Lett. 88 (2001) 017901.

\bibitem {Oppenheim}J. Oppenheim, M. Horodecki, P. Horodecki, and R. Horodecki, Phys. Rev. Lett. 89 (2002) 180402.

\bibitem {Vedral}V. Vedral, Found. Phys. 40 (2010) 1141.


\bibitem {Datta1}A. Datta, A. Shaji, and C. M. Caves, Phys. Rev. Lett. 100 (2008) 050502; B. P. Lanyon, M. Barbieri, M. P. Almeida, and A. G. White, Phys. Rev. Lett. 101 (2008) 200501.

\bibitem {Maziero1}J. Maziero, L. C. C\'eleri, R. M. Serra, and V. Vedral, Phys. Rev. A 80 (2009) 044102.

\bibitem {Maziero2}J. Maziero\textit{ et al.}, Phys. Rev. A 81 (2010) 022116.

\bibitem {Xu}J.-S. Xu \textit{et al.}, Nat. Commun. 1 (2010) 7.

\bibitem {Dillenschneider1}
R. Dillenschneider, Phys. Rev. B 78 (2008) 224413; M. S. Sarandy, Phys. Rev. A 80 (2009) 022108; T. Werlang and G. Rigolin, Phys. Rev. A 81 (2010) 044101; T. Werlang \textit{et al.}, Phys. Rev. Lett. 105 (2010) 09570; Y.-X. Chen and S.-W. Li, Phys. Rev. A 81 (2010) 032120; Y.-X. Chen and Z. Yin, Commun. Theor. Phys. 54 (2010) 60; J. Maziero \textit{et al.}, Phys. Rev. A 82 (2010) 012106; J. Maziero \textit{et al.}, arXiv:1012.5926; B. Tomasello \textit{et al.}, arXiv:1012.4270.

\bibitem {Rosario}
C. A. Rodr\'iguez-Rosario \textit{et al.}, J. Phys. A: Math. Gen. 41 (2008) 205301; A. Shabani and D. A. Lidar, Phys. Rev. Lett. 102 (2009) 100402; T. Werlang, S. Souza, F. F. Fanchini, and C. J. Villas Boas, Phys. Rev. A 80 (2009) 024103; B. Wang, Z.-Y. Xu, Z.-Q. Chen, and M. Feng, Phys. Rev. A 81 (2010) 014101; L. Mazzola, J. Piilo, and S. Maniscalco, Phys. Rev. Lett. 104 (2010) 200401.

\bibitem {Bradler} K. Br\'adler, M. M. Wilde, S. Vinjanampathy, and D. B. Uskov, Phys. Rev. A 82  (2010) 062310.

\bibitem {Horodecki2} M. Horodecki \textit{et al.}, Phys. Rev. A 71 (2005) 062307; W. H. Zurek, Phys. Rev. A 67 (2003) 012320; R. Dillenschneider and E. Lutz, Europhys. Lett. 88 (2009) 50003; A. Brodutch and D. R. Terno, Phys. Rev. A 81 (2010) 062103.

\bibitem {Datta2}A. Datta, Phys. Rev. A 80 (2009) 052304; L. C. C\'eleri \textit{et al.}, Phys. Rev. A 81 (2010) 062130.

\bibitem {Hamieh}S. Hamieh, R. Kobes, and H. Zaraket, Phys. Rev. A 70 (2004) 052325.

\bibitem {Luo1}S. Luo, Phys. Rev. A 77 (2008) 042303.

\bibitem {Modi} K. Modi \textit{et al.}, Phys. Rev. Lett. 104 (2010) 080501.

\bibitem {Terhal}B. M. Terhal \textit{et al.}, J. Math. Phys. 43 (2002) 4286; D. P. DiVincenzo \textit{et al.}, Phys. Rev. Lett. 92 (2004) 067902; S. Luo, Phys. Rev. A 77 (2008) 022301; M. Piani, M. Christandl, C. E. Mora, and P. Horodecki, Phys. Rev. Lett. 102 (2009) 250503; M. H. Partovi, Phys. Rev. Lett. 103 (2009) 230502.

\bibitem{Piani} M. Piani, P. Horodecki, and R. Horodecki, Phys. Rev. Lett. 100 (2008) 090502.

\bibitem {NMR}D. O. Soares-Pinto \textit{et al.}, Phys. Rev. A 81 (2010) 062118; R. Auccaise \textit{et al.}, arXiv:1104.1596.

\bibitem {Nielsen} M. A. Nielsen and I. L. Chuang, \textit{Quantum Computation and Quantum Information} (Cambridge University Press, Cambridge, 2000).

\bibitem{Petz} P. Hayden, R. Jozsa, D. Petz, and A. Winter, Commun. Math. Phys. 246 (2004) 359.

\bibitem {Ali} M. Ali, A. R. P. Rau, and G. Alber, Phys. Rev. A 81 (2010) 042105.


\end{thebibliography}

\end{document}